# Development, validation and clinical usefulness of a prognostic model for relapse in relapsing-remitting multiple sclerosis

Konstantina Chalkou[1], Ewout Steyerberg[2], Patrick Bossuyt[3], Suvitha Subramanian[4], Pascal Benkert[4], Jens Kuhle[5,6], Giulio Disanto[7], Ludwig Kappos[5], Chiara Zecca[8,9], Matthias Egger[1], Georgia Salanti[1]

**Affiliation:** Institute of Social and Preventive Medicine, University of Bern, Bern, **Affiliation:** Institute of Social and Preventive Medicine, University of Bern, Bern, Switzerland [1]; Leiden University Medical Center, Leiden, the Netherlands [2]; Department Epidemiology and Data Science , Amsterdam University Medical Centres, University of Amsterdam, Amsterdam, the Netherlands [3]; Department of clinical research, University of Basel, Basel, Switzerland[4]; Multiple Sclerosis Centre, Neurologic Clinic and Policlinic, Departments of Head, Spine and Neuromedicine, Biomedicine and Clinical Research, University Hospital Basel and University of Basel, Basel, Switzerland[5]; Research Center for Clinical Neuroimmunology and Neuroscience (RC2NB), University Hospital and University of Basel, Switzerland[6]; Neurocenter of Southern Switzerland, Civic Hospital, Lugano, Switzerland[7]; Multiple Sclerosis Center, Neurocenter of Southern Switzerland, EOC, Lugano, Switzerland [8]; Faculty of biomedical Sciences, Università della Svizzera Italiana, Lugano, Switzerland [9]

Correspondence to:

Konstantina Chalkou, MSc

Institute of Social & Preventive Medicine

University of Bern

Mittelstrasse 43, 3012 Bern, Switzerland

konstantina.chalkou@ispm.unibe.ch

***Abstract***

*Background:* Prognosis for the occurrence of relapses in individuals with Relapsing-Remitting Multiple Sclerosis (RRMS), the most common subtype of Multiple Sclerosis (MS), could support individualized decisions and disease management and could be helpful for efficiently selecting patients for future randomized clinical trials. There are only three previously published prognostic models on this, all of them with important methodological shortcomings.

*Objectives:* We aim to present the development, internal validation, and evaluation of the potential clinical benefit of a prognostic model for relapses for individuals with RRMS using real world data.

*Methods:* We followed seven steps to develop and validate the prognostic model; 1) selection of prognostic factors via a review of the literature, 2) development of a generalized linear mixed effects model in a Bayesian framework, 3) examination of sample size efficiency, 4) shrinkage of the coefficients, 5) dealing with missing data using multiple imputations, 6) internal validation of the model. Finally, we evaluated the potential clinical benefit of the developed prognostic model using decision curve analysis. For the development and the validation of our prognostic model, we followed the TRIPOD statement.

*Results:* We selected eight baseline prognostic factors: age, sex, prior MS treatment, months since last relapse, disease duration, number of prior relapses, expanded disability status scale (EDSS) score, and number of gadolinium enhanced lesions. We also developed a web application that calculates an individual's probability of relapsing within two years. The optimism-corrected c-statistic is 0.65 and the optimism-corrected calibration slope is 0.92.

For threshold probabilities between 15% and 30%, the "treat based on the prognostic model" strategy leads to the highest net benefit and hence is considered the most clinically useful strategy.

*Conclusions:* The prognostic model we developed offers several advantages in comparison to previously published prognostic models on RRMS. Importantly, we assessed the potential clinical benefit to better quantify the clinical impact of the model. Our web application, once externally validated in the future, could be used by patients and doctors to calculate the individualized probability of relapsing within two years and to inform the management of their disease.

## *1 Introduction*

Multiple sclerosis (MS) is an immune-mediated disease of the central nervous system with several subtypes. The most common subtype is relapsing-remitting multiple sclerosis (RRMS).[1] Patients with RRMS present with acute or subacute symptoms (relapses) followed by periods of complete or incomplete recovery (remissions).[2] Effective treatment of patients with RRMS can prevent disease progression and associated severe consequences, like spasticity, fatigue, cognitive dysfunction, depression, bladder dysfunction, bowel dysfunction, sexual dysfunction, pain and death.[3]

Relapses have been commonly used as a primary efficacy endpoint in phase III randomized clinical trials leading to market approval of RRMS therapies, although the strength of the association between relapses and disease progression (an outcome of highest interest to patients) is still debated.[4, 5, 6] Prognosis for relapses in individuals with RRMS could support individualized decisions and disease management. A prognostic model for relapses may also be helpful for the efficient selection of patients in future randomized clinical trials and, therefore, for the reduction of type II errors in these trials.[7] In addition, such a model could support individualized decisions on initiation or switch of disease modifying treatment (DMT). To our knowledge, no widely accepted prognostic model for MS has been used in clinical practice yet.

A recent systematic review of prediction models in RRMS [8] identified only three prognostic models (i.e. models that focus on predicting the outcome instead of predicting treatment response) with relapses as the outcome of interest.[7, 9, 10] However, all three studies had methodological shortcomings. Only one small study, with 127 patients, used a cohort of patients that is considered the best source of prognostic information. [8] All three studies used complete cases, excluding cases with missing data, analysis without justifying the assumptions underlying this approach; given the potential non-random distribution of missing data the results might be biased [11]. In addition, none of them validated internally their model and they did not present calibration or discrimination measures. Hence, they might be at risk of misspecification.[12] In addition, none of them used shrinkage to avoid overfitted models.[13] Finally, none of the studies evaluated the clinical benefit of the model, an essential step, which quantifies whether and to what extent a prognostic model is potentially useful in decision-making and clinical practice. Similar limitations exist in other published prognostic models, which commonly have serious deficiencies in the statistical methods, are based on small datasets, and have inappropriate handling of missing data and lack validation.[14]

In this research work, we aim to fill the gap of prognostic models on relapses for RRMS patients. We present the development, the internal validation, and the evaluation of the clinical benefit of a prognostic model for relapses for individuals with RRMS using real world data from the Swiss Multiple Sclerosis Cohort (SMSC).[15] The cohort is comprised of patients diagnosed with RRMS who are followed bi-annually or annually in major MS Centres with full standardized neurological examinations, MRIs and laboratory investigations.[15] Our prognostic model is designed for a patient who, within the Swiss health care system and standard MS treatment protocols, would like to estimate their probability of having at least one relapse within the next two years.

## *2 Data and Methods*

In section 2.1 we describe the data available for the model development. We followed seven steps (described in detail in section 2.3) to build and evaluate the prognostic model; 1) selection of prognostic factors via a review of the literature, 2) development of a generalized linear mixed effects model in a Bayesian framework, 3) examination of sample size efficiency, 4) shrinkage of the coefficients, 5) dealing with missing data using multiple imputations, 6) internal validation of the model. Finally, we evaluated the potential clinical benefit of the developed prognostic model. For the development and the validation of our prognostic model we followed the TRIPOD statement;[16] the TRIPOD checklist is presented in the **Appendix Table 1**.

### 2.1 Data Description

We analyzed observational data on patients diagnosed with relapsing-remitting multiple sclerosis (RRMS) provided by the Swiss Multiple Sclerosis Cohort (SMSC)) study,[15] which has been recruiting patients since June 2012. SMSC is a prospective multicenter cohort study performed across seven Swiss centers. Every patient included in the cohort is followed-up every 6 or 12 months, and the occurrence of relapses, disability progression, DMTs initiation or interruption, adverse events, and concomitant medications are recorded at each visit. Brain MRI and serum samples are also collected at each visit. The strength of SMSC is the high quality of data collected including MRI scans and body fluid samples in a large group of patients. In addition, several internal controls and validation procedures are performed to ensure the quality of the data.

We included patients with at least two years follow-up. The drop-out rate in the entire SMSC cohort was 15.8%. Drop-out was primarily associated with change of address and health care provided by a physician not associated with SMSC. Therefore, we assume that patients dropping out of the cohort before completing two years were not more likely to have relapsed than those remaining in the cohort, and hence the risk of attrition bias is low. The dataset includes 935 patients, and each patient has one, two or three two-year follow-up cycles. At the end of each two-year cycle, we measured relapse occurrence as a dichotomous outcome. At the beginning of each cycle, several patient characteristics are measured and we considered them as baseline characteristics for this specific cycle. In total, we included 1752 cycles from the 935 study participants. Patients could be prescribed several potential DMTs during their follow-up period, i.e., a patient during a two-year follow-up cycle could either take no DMT or one of the available DMTs. We used the treatment status only at baseline of each two-year cycle to define the dichotomous prognostic factor 'currently on treatment' or not.

We transformed some of the continuous variables to better approximate normal distributions and merged categories with very low frequencies in categorical variables. **Table *1*** presents summary statistics of some important baseline characteristic using all cycles (n=1752), while in **Appendix Table 2**, we present the outcome of interest (frequency of relapse within two years), as well as several baseline characteristics separately for patients that were included in 1 cycle, patients that were included in 2 cycles, and patients that were included in 3 cycles.

**Table 1 summary statistics of some important baseline characteristic using all 1752 two-years cycles coming from 935 unique patient in SMSC**

| **Characteristics** | **Number of observations (n=1752)** |
|---|---|
| **Relapse within two years** | |
| *Yes* | |
| n (%) | 302 (17.2) |
| *No* | |
| n (%) | 1450 (82.8) |
| **Gender** | |
| *Females* | |
| n (%) | 1209 (69) |
| *Males* | |
| n (%) | 543 (31) |
| **Currently on treatment** | |
| *Yes* | 1639 (93.6) |
| n (%) | |
| *No* | |
| n (%) | 113 (6.4) |
| *NA* | |
| n (%) | 34 (2.0) |
| **Age** | |
| mean ± sd | 42.4 ± 11.3 |
| min | 18 |
| max | 76.4 |
| **Disease Duration (years)** | |
| mean ± sd | 10.9 ± 8.3 |
| min | 0.0 |
| max | 41.2 |
| **EDSS** | |
| mean ± sd | 2.4 ± 1.4 |
| min | 0.0 |
| max | 7.0 |
| **Number of Gadolinium enhanced lesions** | |
| *=0* | |
| n (%) | 956 (55.0) |
| *=1* | |
| n (%) | 26 (1.0) |
| *≥ 2* | |
| n (%) | 25 (1.0) |
| *NA* | |
| n (%) | 745 (43.0) |

**2.2 Notation**

Let $Y_{ij}$ denote the dichotomous outcome for individual $i$ where $i$=1, 2, ..., $n$ at the $j^{th}$ two-year follow-up cycle out of $c_i$ cycles. $PF_{ijk}$ is the $k^{th}$ prognostic factor k=1,…, $np$. An individual develops the outcome $(Y_{ij} = 1)$ or not $(Y_{ij} = 0)$ according to its probability $p_{ij}$.

## 2.3 Steps in building the prognostic model

### *2.3.1 Step 1 - Selection of prognostic factors*

Developing a model using a set of predictors informed by prior knowledge (either in the form of expert opinion or previously identified variables in other prognostic studies) has conceptual and computational advantages.[17, 18, 19] Hence, in addition to the information obtained from the three prognostic models included in the recent systematic review discussed in introduction [7, 9, 10,] we aimed to increase our relevant information, via searching for prediction models or research works aiming to identify subgroups of patients in RMMS. We searched in PubMed (https://pubmed.ncbi.nlm.nih.gov), using the string *((((predict*[Title/Abstract] OR prognos*[Title/Abstract])) AND Relapsing Remitting Multiple Sclerosis[Title/Abstract]) AND relaps*[Title/Abstract]) AND model[Title/Abstract].* We then decided to build a model with all prognostic factors included in at least two of the previously published models.

### *2.3.2 Step 2 - Logistic mixed-effects model*

We developed a logistic mixed-effects model in a Bayesian framework:

*Model 1*

$$Y_{ij} \sim Bernoulli(p_{ij})$$

$$logit(p_{ij}) = \beta_0 + u_{oi} + \sum_{k=1}^{np} (\beta_k + u_{ki}) \times PF_{i,k,j}$$

We used fixed effect intercept ($\beta_0$), fixed effect slopes ($\beta_k$), individual-level random effects intercept ($u_{oi}$), and individual-level random effects slopes ($u_{ki}$) to account for information about the same patient from different cycles.

We define $\boldsymbol{u} = \begin{pmatrix} \boldsymbol{u_o} \\ \boldsymbol{u_k} \end{pmatrix}$ to be the $(np + 1) \times n$ matrix of all random parameters and we assume it is normally distributed $\boldsymbol{u} \sim \mathrm{N}(\boldsymbol{0}, \boldsymbol{D_u})$ with mean zero and a $(np + 1) \times (np + 1)$ variance-covariance matrix

$$D_u = \begin{bmatrix} \sigma^2 & \rho \times \sigma^2 & \dots & \rho \times \sigma^2 & \rho \times \sigma^2 \\ \rho \times \sigma^2 & & & & \rho \times \sigma^2 \\ \vdots & & \ddots & & \vdots \\ \rho \times \sigma^2 & & & & \rho \times \sigma^2 \\ \rho \times \sigma^2 & \rho \times \sigma^2 & \dots & \rho \times \sigma^2 & \sigma^2 \end{bmatrix}$$

This structure assumes that the variances of the impact of the variables on multiple observations for the same individual are equal ($\sigma^2$) and that the covariances between the effects of the variables are equal too ($\rho \times \sigma^2$).

*2.3.3 Step 3 - Examination of sample size efficiency*

We examined if the available sample size was enough for the development of a prognostic model.[16] We calculated the events per variable (EPV) accounting for both fixed-effects and random-effects and for categorical variables.[20] We also used the method by Riley et al. to calculate the efficient sample size for the development of a logistic regression model, using the R package `pmsampsize`.[21] We set Nagelkerke's R2 = 0.15 (Cox-Snell's adjusted R2 = 0.09) and the desired shrinkage equal to 0.9 as recommended. [21]

*2.3.4 Step 4 - Shrinkage of the coefficients*

The estimated effects of the covariates need some form of penalization to avoid extreme predictions.[13, 22] In a Bayesian setting, recommended shrinkage methods use a prior on the regression coefficients. [23] For logistic regression, a Laplace prior distribution for the regression coefficients is recommended [24] (i.e. double exponential, also called Bayesian LASSO)

$$\pi(\beta_1, \beta_2, \dots, \beta_{np}) = \prod_{k=1}^{np} \frac{\lambda}{2} e^{-\lambda|\beta_k|},$$

where $\lambda$ is the shrinkage parameter. A Laplace prior allows small coefficients to shrink towards zero faster, while it applies smaller shrinkage to large coefficients.[25]

*2.3.5 Step 5 - Multiple imputations for missing data*

In the case of missing values in the covariates, we assumed that these are missing at random (MAR), meaning that, given the observed data, the occurrence of missing values is independent of the actual missing values. Appropriate multiple imputation models should provide valid and efficient estimates if data are MAR. As our substantive model is hierarchical, we used Multilevel Joint Modelling Multiple imputations using the `mitml` R package.[26]

First, we checked for variables not included in the substantive model that could predict the missing values (i.e. auxiliary variables). Then we built the imputation model, using both fixed effect and individual-level random effects intercept and slopes as in our substantive (Model 1), where the dependent variables are the variables that include missing values for imputation, and the independent variables are all complete variables included in the substantive model and the identified auxiliary variables.

We generated 10 imputed datasets, using the `jomoImpute` R function, and we applied the Bayesian model (Model 1) to each of the imputed datasets. We checked convergence of the imputations using the plot R function in the `mitml` R package. Finally, we obtained the pooled estimates for the regression coefficients, $\widehat{\beta_0}$ and $\widehat{\beta_k}$, using Rubin's rules [27] (`testEstimates` R function) with two matrices containing the mean and the variances estimates, respectively, from each imputed dataset as arguments.

*2.3.6 Step 6 - Internal Validation*

First, we assessed the calibration ability of the developed model, via a calibration plot with loess smoother, for the agreement between the estimated probabilities of the outcome and the observed outcome's proportion (`val.prob.ci.2`) R function. We used bootstrap internal validation to correct for optimism in the calibration slope and in discrimination, measured via the AUC,.[13] For each one of the 10 imputed datasets, we created 500 bootstrap samples and in each one of them: 1) we constructed a generalized linear model with the pre-specified predictors, using `glm` R function, denoted as Model*, 2) we calculated the bootstrap performance as the apparent performance of Model* on the sample for each one of the bootstrap samples, 3) we applied the Model* to the corresponding imputed dataset to determine the test performance, 4) we calculated the optimism as the difference between bootstrap performance and test performance. Then we calculated the average optimism between the 500 bootstrap samples and used Rubin's rules to summarize the optimism for the AUC and the calibration slope between the 10 imputed datasets. We calculated the optimism-corrected AUC and calibration slope of our prognostic model, by subtracting the optimism estimate from the apparent performance.

Ideally, we should construct the Bayesian logistic mixed effects model exactly as we developed the original model. However, this would need 15000 hours to run, as the Bayesian model needs to run for 500 bootstrap samples in each one of the 10 imputed datasets (i.e. 5000 times) and the Bayesian model itself needs 3 hours, and hence the bootstrap internal validation we performed results to a rough optimism estimation ignoring the dependence between the same individual.

We used self-programming R routines to validate the model via bootstrapping.

*2.3.7 Clinical benefit of the developed model*

Decision curve analysis is a widely used method to evaluate the clinical consequences of a prognostic model. This method aims to overcome some weaknesses of the traditional measures (i.e. discrimination and calibration) that are not informative about the clinical value of the prognostic model.[28] Briefly, decision curve analysis calculates a clinical "net benefit" for a prognostic model and compares it in with the default strategies of treat all or treat none of the patients. Net benefit (NB) is calculated across a range of threshold probabilities, defined as the minimum probability of the outcome for which a decision will be made.

More detailed, information about their risk of relapsing within two years might be important to help patients to re-consider whether their current treatment and approach should continue to follow the established standards of care in Switzerland. If the probability of relapsing is considered too high, maybe RRMS patients would be interested on taking a more radical stance towards the management of their condition: discuss with their treating doctors about more active disease-modifying drugs (which might also have high risk of serious adverse events), explore the possibility of stem cell transplantation etc. Let us call this the "more active approach". If the probability of relapsing is higher than a threshold α% then a patient will take a "more active approach" to the management of their condition, otherwise they will continue "as per standard care".

We examined the net benefit of our final model, via the estimated probabilities provided by the equation in Section 2.3.6, by using decision curve analysis and plotting the NB of the developed prognostic model, using the `dca` R function, in a range of threshold probabilities α% that is equal to

$$NB_{decision\ based\ on\ the\ model} = (True\ positive\ \%) - (False\ positive\ \%) \times \frac{\alpha\%}{1-\alpha\%}\ .$$

We compare the results with those from two default strategies: recommend "as per standard care for all" and continue "more active approach for all". The NB of "as per standard care for all" is equal to zero in the whole range of the threshold probabilities, as there are no false positives and false negatives. "More active approach for all" does not imply that the threshold probability $a\%$ has been set to 0 and is calculated for the whole range of threshold probabilities using the formula:

$$NB_{more\ active\ approach\ for\ all} = (prevalence) - (1 - prevalence) \times \frac{a\%}{1-a\%}$$

These two strategies mean the more active treatment options will be discussed and considered by all patients ("more active approach for all") or with none ("as per standard care for all"). A decision based on a prognostic model is only clinically useful at threshold $a\%$ if it has a higher NB than both "more active approach for all" and ("as per standard care for all"). If a prognostic model has a lower NB than any default strategy, the model is considered clinically harmful, as one of the default strategies leads to better decisions.[28, 29, 30, 31, 32]

We made the analysis code available in a GitHub library: https://github.com/htx-r/Reproduce-results-from-papers/tree/master/PrognosticModelRRMS

### *3 Results*

For the model development, we used 1752 observations coming from two-years repeated cycles of 935 patients who experienced 302 relapses

First, we took into account the three prognostic models included in the recent systematic review [7, 9, 10] that predict relapse (not the treatment response to relapses) in patients with RRMS. Our search in PubMed identified 87 research articles. After reading the abstracts, we ended up with seven models that predicted either relapses or treatment response to relapses.

Three of them were already included in the recent systematic review, as they predicted relapses. Hence, we identified three additional models that predict the treatment response to relapses [33, 34, 35,] and one research work aiming to identify subgroups of RRMS patients who are more responsive to treatments.[36]

***Figure 1*** shows which prognostic factors were selected and which pre-existing prognostic models were included. [7, 9, 33, 34, 35, 36] We included none of the prognostic factors included in Liguori et. al.[10] model, as none of the prognostic factors they used (i.e. MRI predictors) were included in any other of the available models. We briefly summarize these models in Section 1 of the Appendix file, and some important characteristics of these models are shown in **Appendix Table 3**.

***Figure 1 Venn Diagram of the prognostic factors included at least two times in pre-existing models and included in our prognostic model. The names with * refer to the first author of each prognostic model or prognostic factor research.*** [7, 9, 10, 33, 34, 35, 36]

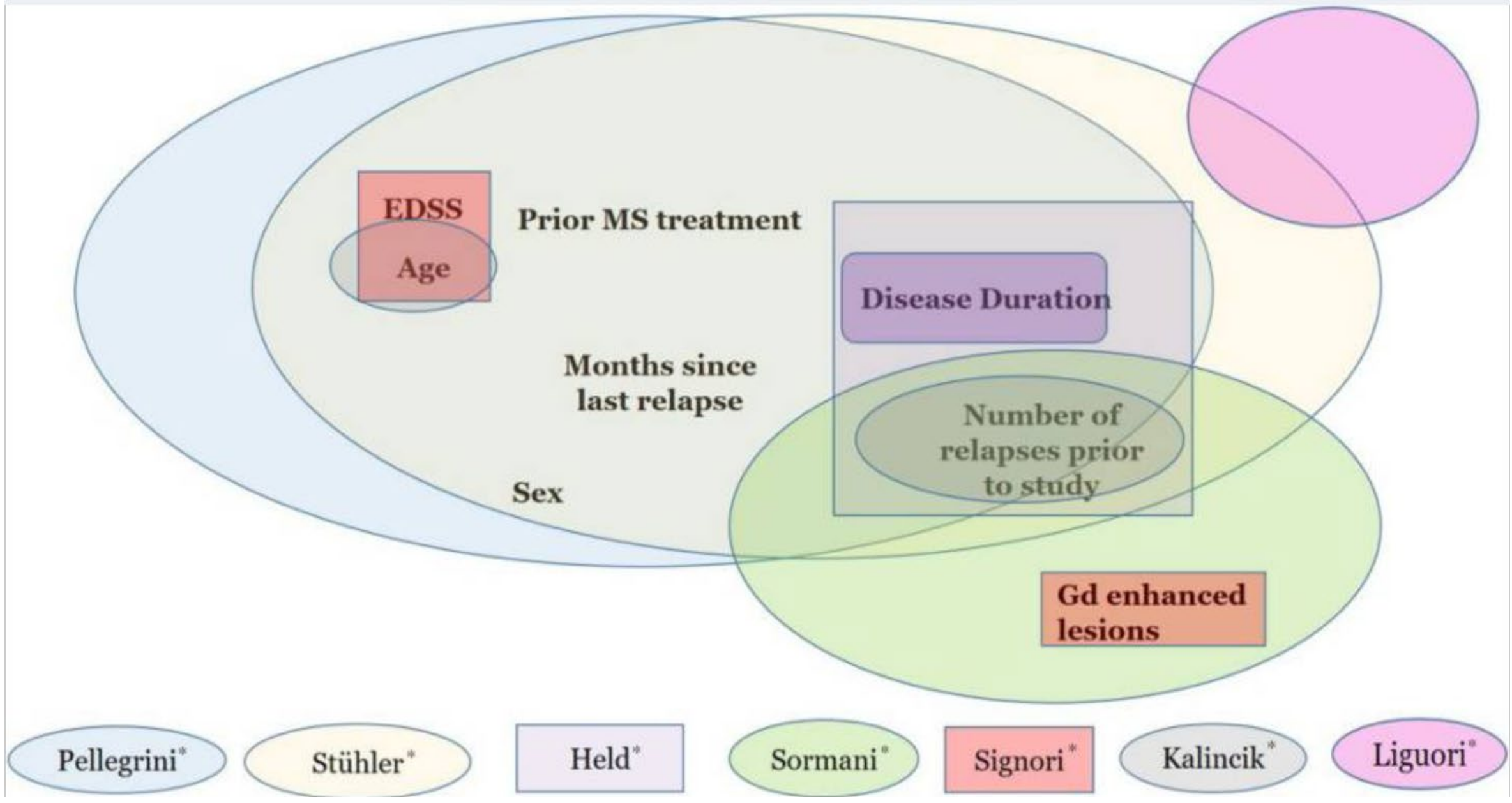

***EDSS: Expanded Disability Status Scale, Gd: Gadolinium***

prognostic factors included in our model are presented in ***Table 2*** with their pooled estimated $\widehat{\beta_k}$, ORs and their corresponding 95% Credible Intervals (CrIs). We have also developed a web application where the personalized probabilities to relapse within two years are calculated automatically. This is available for use in a R Shiny app https://cinema.ispm.unibe.ch/shinies/rrms/. In this example the variance $\sigma^2$ is estimated 0.0001 and the covariance $\rho \times \sigma^2$ are equal to 0.00005. Hence, the random intercept and all random slopes were estimated close to 0. For convenience and speed of estimation, predictions were made using only the fixed effects estimates. In the Supplementary file, ***Appendix Table 4***, we present the estimated coefficients in each of the ten imputed datasets.

**Table 2 Pooled estimates of the regression coefficients $\widehat{\beta_k}$, ORs and the 95% CrIs for each one of the parameters in the model (centralized to the mean), using the Rubin's rules. The estimated $\sigma$ (standard deviation of the impact of the variables on multiple observations for the same individuals) is 0.01. The estimated correlation $\rho$ between the effects of the variables is 0.49. The pooled optimism-corrected AUC is 0.65 and the pooled optimism-corrected calibration slope is 0.91.**

| Parameters | $\widehat{\beta_k}$ | *OR* | *95% CrI* |
|---|---|---|---|
| Age | -0.035 | 0.97 | (0.95, 0.98) |
| Disease duration | 0.337 | 1.40 | (0.90, 2.18) |
| EDSS | 0.122 | 1.13 | (1.02, 1.25) |
| Number of gadolinium enhanced lesions (>0 vs 0) | -0.034 | 0.97 | (0.69, 1.36) |
| Number of previous relapses (1 vs 0) | -0.070 | 0.93 | (0.69, 1.26) |
| Number of previous relapses (2 or more vs 0) | 0.133 | 1.14 | (0.81, 1.61) |
| Months since last relapse | -0.478 | 0.62 | (0.49, 0.78) |
| Treatment naïve (Yes vs No) | 0.086 | 1.09 | (0.80, 1.49) |
| Gender (Female vs Male) | 0.254 | 1.29 | (0.97, 1.72) |
| On treatment (Yes vs No) | -0.221 | 0.80 | (0.50, 1.27) |

**Disease duration is transformed to $log(disease\ duration + 10)$**

**Months since last relapse is transformed to $log(months\ since\ last\ relapse + 10)$**

The full model's degrees of freedom were 22 (for 10 predictors with random intercept and slope) and the events per variable (EPV) was 13.7. The efficient sample size was calculated as 2084 (to avoid optimism in the regression coefficients), 687 (for agreement between apparent and adjusted model performance), and 220 (for a precise estimation of risk in the whole population).[21] Our available sample size suggests that there might be optimism in our regression coefficients. However, this should have been addressed via the shrinkage we performed.

In ***Figure 2*** we show the distributions of the calculated probability of relapsing for individuals by relapse status. The overlap in the distributions of the probabilities is large, as also shown by the optimism-corrected AUC (***Table 2***). The overall mean probability of relapsing is 19.1%. For patients who relapsed the corresponding mean is 23.4% whereas for patients who did not relapse is 18.0%. **Figure 3** shows the calibration plot, with some apparent performance measures and their 95% Confidence Intervals (CIs), of the developed prognostic models and represents the agreement between the estimated probabilities and the observed proportion to relapse within two years.

***Figure 2 The distribution of probability of relapsing within two years by relapse status at the end of two-years follow-up cycles. The dashed lines indicate the mean of estimated probability/risk for cycles that ended up with relapse (purple) and for those without relapse (yellow).***

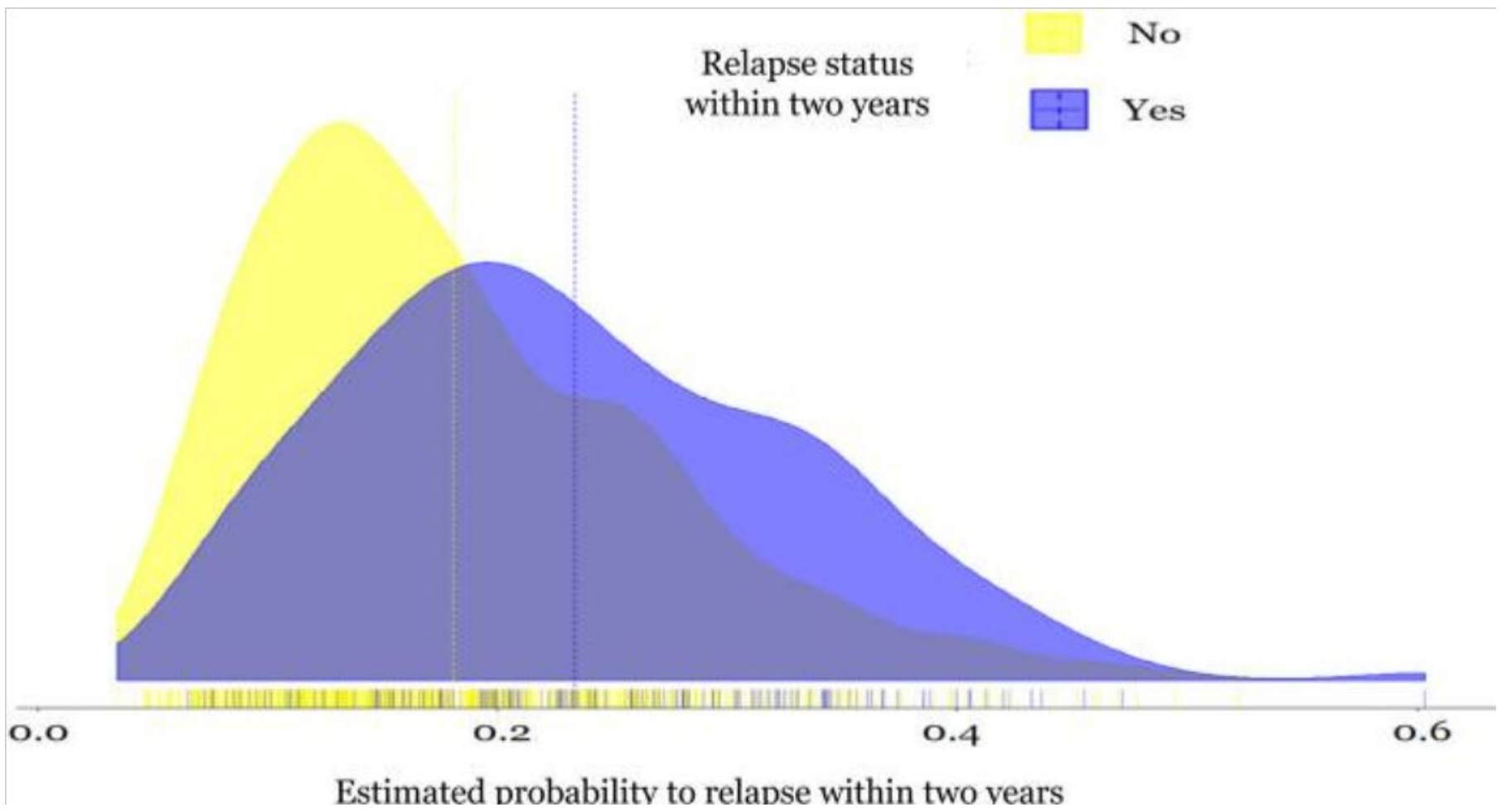

***Figure 3 Calibration plot (N=1752) of the developed prognostic model with loess smoother. The distribution of the estimated probabilities is shown at the bottom of the graph, by status relapse within two years (i.e. events and non-events). The horizontal axis represents the expected probability of relapsing within 2 years and the vertical axis represents the observed proportion of relapse. The apparent performance measures (c-statistic and c-slope) with their correspondent 95% CI are also shown in the graph.***

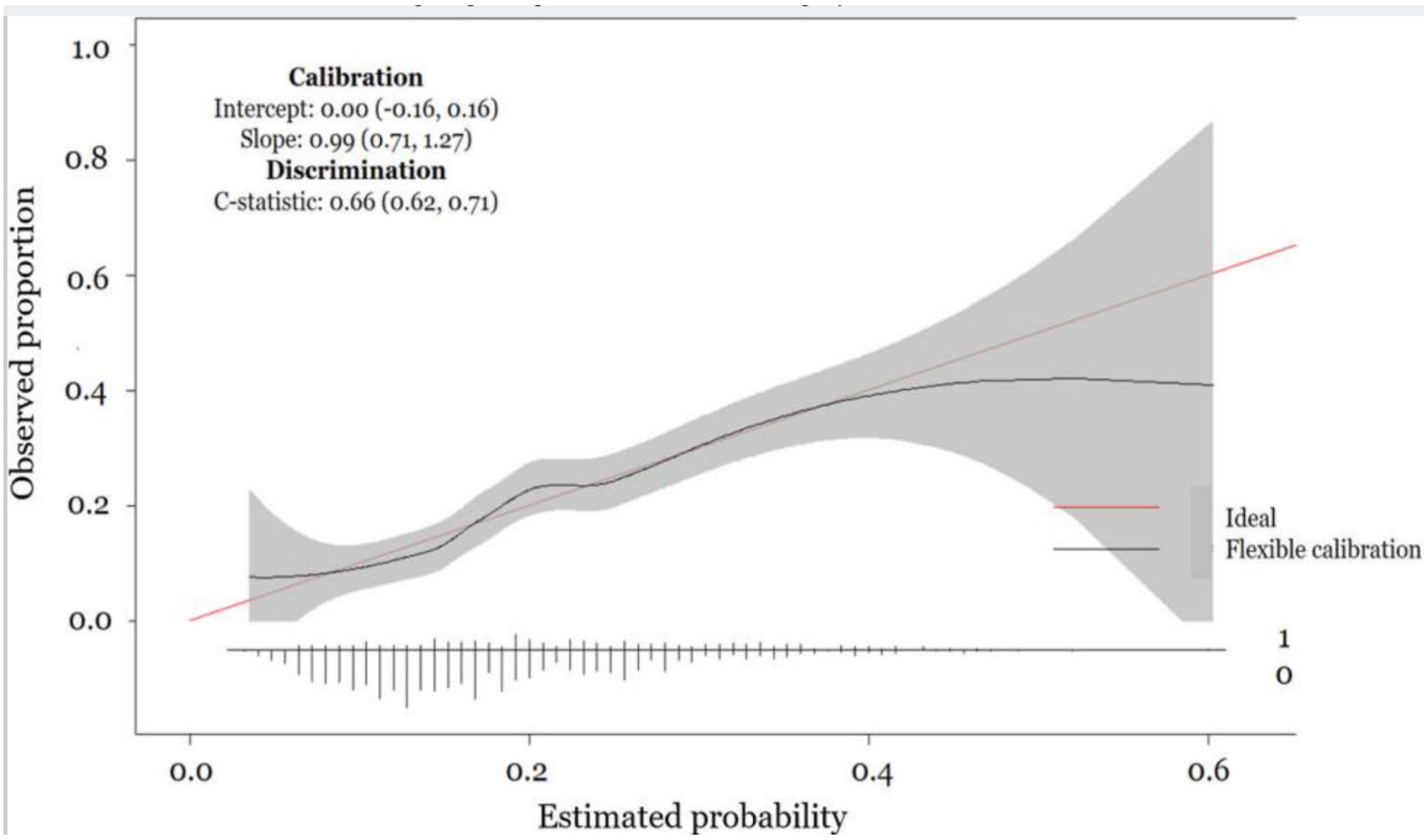

In ***Figure 4,*** the exploration of net benefit of our prognostic model is presented. [29, 30, 31, 32] In the figure, the vertical axis corresponds to the NB and the horizontal axis correspond to the preferences presented as threshold probabilities. The NB is a weight between the benefit of identifying, and consequently correctly treating, individuals that relapsed and the harm (e.g., side effects) of wrongly prescribing patients the “more active approach” due to false positives results. Threshold probabilities refer to how decision makers value the risk of relapsing related to a harmful condition for a given patient, a decision that is often influenced by a discussion between the decision maker and the patient. It is easily seen that the dashed line, corresponding to decisions based on the developed prognostic model, has the highest NB compared to default strategies, between the range 15% and 30% of the threshold probabilities.

Nearly half of the patients (46.5%) in our dataset have calculated probabilities between these ranges, in at least one follow-up cycle. Hence, for patients that consider the relapse occurrence to be 3.3 to 6.6 times worse ($\frac{1}{a\%}$) than the risks, costs and inconvenience in "more active approach", the prognostic model can lead to better decisions than the default strategies.

***Figure 4 Decision curve analysis showing the net benefit of the prognostic model per cycle. The horizontal axis is the threshold estimated probability of relapsing within two years, $a\%$, and the vertical axis is the net benefit. The plot compares the clinical benefit of three approaches: "as per standard care for all" approach, "more active care for all" approach, and "decision based on the prognostic model" approach (see definitions on Section 2.3.8.). For a given threshold probability, the approach with the highest net benefit is considered the most clinically useful model. The "decision based on the prognostic model" approach provides the highest net benefit for threshold probabilities ranging from 15%-30%.***

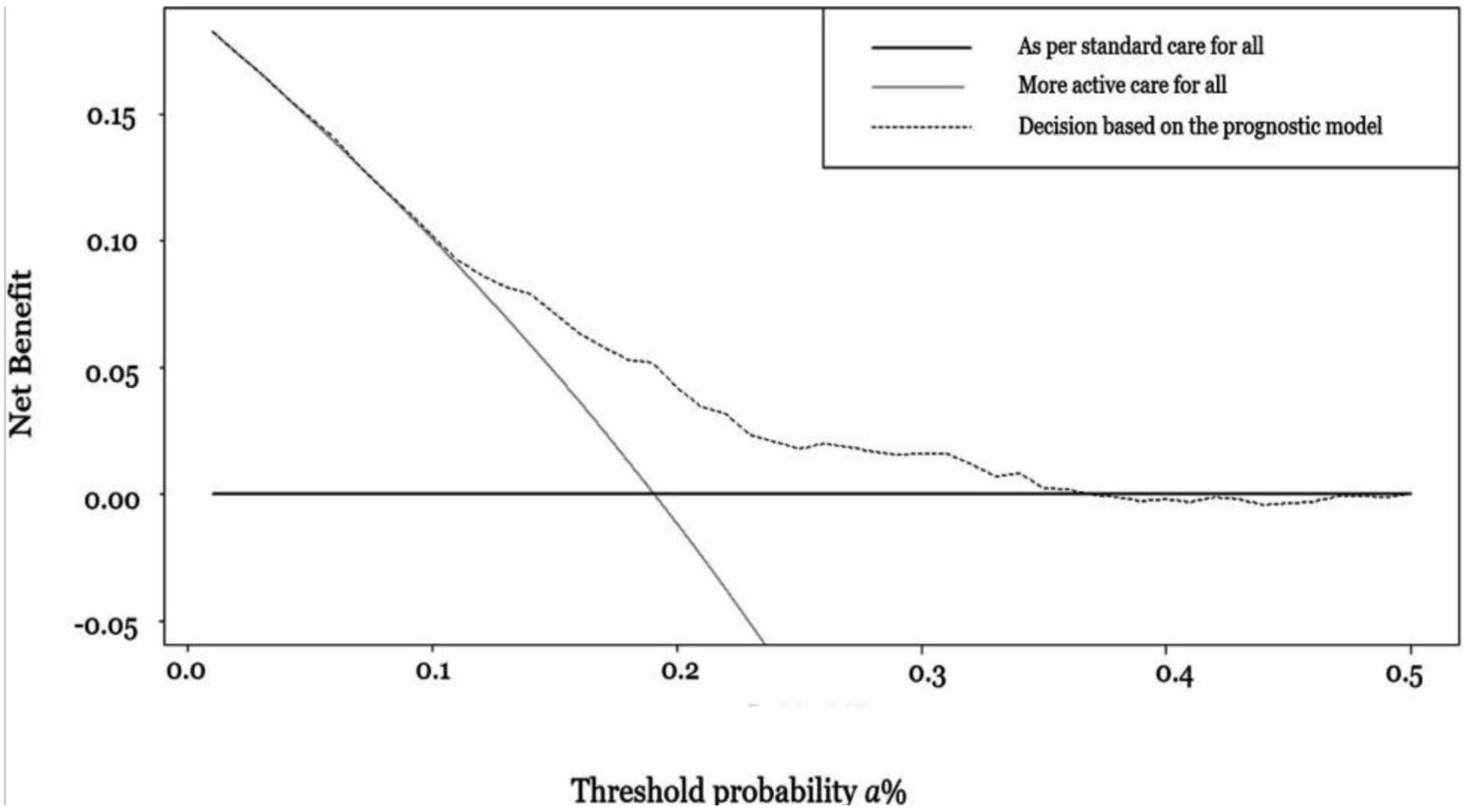

## *4 Discussion*

We developed a prognostic model that predicts relapse within two years for individuals diagnosed with RRMS, using observational data from the SMSC, [15] a prospective multicenter cohort study, to inform clinical decisions. Prognostication is essential for the disease management of RRMS patients and until now, no widely accepted prognostic model for MS is used in clinical practice. A recent systematic review on prognostic models for RRMS,[8] describes that most of the prognostic models, regardless the outcome of interest, are lacking statistical quality in the development steps, introducing potential bias, did not perform internal validation, did not report important performance measures like calibration and discrimination and did not present the clinical impact of the models. More specifically, only three studies examined the relapses as an outcome of interest and none of them satisfied the criteria above. Our model aimed to fill the existing gap, by satisfying all the above criteria, to enhance the available information for predicting relapses and to inform decision-making.

Given that a manageable number of characteristics is needed to establish the risk score, doctors and patients can enter these using our online tool (https://cinema.ispm.unibe.ch/shinies/rrms/), estimate the probability of relapsing within two years and take treatment decisions based on patient's risk score. This tool shows the potential of the proposed approach, however, may not yet be ready for use in clinical practice, as decision-making tools need external validation with an independent cohort of patients.

We included eight prognostic factors (all measured at baseline where also the risk was estimated): Age, disease duration, EDSS, number of gadolinium enhanced lesions, number of previous relapses two years prior, months since last relapse, treatment naïve, gender and 'currently on treatment'. The EPV of our model is 13.7, the sample size is efficient enough, and more than the sample size of all three pre-existing prognostic models. The optimism corrected AUC of our model is 0.65, indicating a relatively small discrimination ability of the model. However, in the literature only Stühler et. al. reported the AUC of their model that was also equal to 0.65. In our previous work,[37] the optimism corrected AUC using the LASSO model, with many candidate predictors, was 0.60, whereas this of the pre-specified model was 0.62. This could indicate that, in general, relapses are associated with unknown factors. The prognostic model we developed seems to be potentially useful, preferred over "Treat all" or "Treat none" approaches for threshold ranges between 15% and 30%.

The applicability of our model is limited by several factors. First, the risk of relapsing is not the only outcome that patients will consider when making decisions; long-term disability status would also determine their choice,[4] and there is an ongoing debate of whether the relapse rate is associated with the long-term disability. [5, 6, 7, 38] That could be a further line of future research, and prognostic models with good statistical quality for long-term disability still need to be developed. In addition, the sample size of the SMSC is relatively small compared to other observational studies; this study though is of high quality. Furthermore, the bootstrap internal validation we performed ignores the dependence between the same individuals. In each one of the 10 imputed datasets and the 500 bootstrap samples we constructed a frequentist logistic linear model. Ideally, we should construct the Bayesian logistic mixed effects model exactly as we developed the original model. In addition, for model parsimony reasons, our model assumes that the variances of the impact of the variables on multiple observations for the same individual are equal and that the covariances between the effects of the variables are equal too. This assumption might be relaxed, by e.g., assuming covariate-specific correlations. Finally, our model was not validated externally, something essential for decision-making tools. In the near future, independent researchers, as recommended by Colins et. al.,[39] should validate externally our model before it is ready for clinical use.

## *5 Conclusions*

The prognostic model we developed offers several advantages in comparison to previously published prognostic models in RRMS. We performed multiple imputations for the missing data to avoid potential bias induced,[11] we used shrinkage of the coefficients to avoid overfitting,[13] and we validated internally our model presenting calibration and discrimination measures, an essential step in prognosis research.[13] Importantly, we assessed the net benefit of our prognostic model, which helps to quantify the potential clinical impact of the model. Our web application, when externally validated, could be used by patients and doctors to calculate the individualized risk of relapsing within two years and to inform their decision-making.

**Ethical Approval and Consent to participate**

The use of data for this study was approved by the Cantonal Ethics commission of Bern (Kantonale Ethikkommission für die Forschung, KEK Bern) for the project with ID 2019-02151

**Consent for publication**

The manuscript does not contain any individual person's data in any form

**Availability of data and materials**

The data that support the findings of this study were available from Swiss Multiple Sclerosis Cohort (SMSC). Restrictions apply to the availability of these data, which were used under license for this study.

**Competing interest**

KC, ES, PB, SS, PB, JK, GD, ME, and GS declare that they have no conflict of interest with respect to this paper. LK´s institution (University Hospital Basel) has received in the last 3 years and used exclusively for research support: steering committee, advisory board, and consultancy fees (Actelion [Janssen/J&J], Addex, Bayer, Biogen, Biotica, Genzyme, Lilly, Merck, Novartis, Receptos, Sanofi, Santhera, Siemens, Teva, UCB, and Xenoport); speaker fees (Bayer, Biogen, Merck, Novartis, Sanofi, and Teva); support of educational activities (Bayer, Biogen, CSL Behring, Genzyme, Merck, Novartis, Sanofi, and Teva); license fees for Neurostatus products; and grants (Bayer, Biogen, Merck, Novartis, Roche Research Foundation). CZ received honoraria for speaking and/or consulting fees and/or grants from Abbvie, Almirall, Biogen Idec, Celgene, Genzyme, Lilly, Merck, Novartis, Roche, Teva Pharma.

**Funding**

European Union's Horizon 2020 research and innovation program under grant agreement No 825162.

**Authors' contributions**

KC and GS developed the theory. KC performed the analyses, interpreted the results and wrote the manuscript with a great contribution and support from GS. ES, PB and ME supported, commented on, and contributed to the statistical analyses and interpretations. SS, PB, JK, GD, LK and CZ made substantial contributions to acquisition of data, medical conceptions, and writing the manuscript. All authors read and approved the final manuscript.

**Acknowledgements**

KC, and GS are funded by the European Union's Horizon 2020 research and innovation program under grant agreement No 825162. The authors thank Junfeng Wang and Gobbi Claudio for their comments and their assistance on this article.

**References**

1. Ghasemi N, Razavi S, Nikzad E. Multiple Sclerosis: Pathogenesis, Symptoms, Diagnoses and Cell-Based Therapy. Cell J Yakhteh. 2017;19(1):1-10.
2. Goldenberg MM. Multiple Sclerosis Review. Pharm Ther. 2012;37(3):175-184.
3. Crayton HJ, Rossman HS. Managing the symptoms of multiple sclerosis: A multimodal approach. Clin Ther. 2006;28(4):445-460. doi:10.1016/j.clinthera.2006.04.005
4. Lublin FD. Relapses do not matter in relation to long-term disability: no (they do). Mult Scler Houndmills Basingstoke Engl. 2011;17(12):1415-1416. doi:10.1177/1352458511427515

5. Casserly C, Ebers GC. Relapses do not matter in relation to long-term disability: yes. Mult Scler Houndmills Basingstoke Engl. 2011;17(12):1412-1414. doi:10.1177/1352458511427514
6. Hutchinson M. Relapses do not matter in relation to long-term disability: commentary. Mult Scler Houndmills Basingstoke Engl. 2011;17(12):1417. doi:10.1177/1352458511427512
7. Sormani MP, Rovaris M, Comi G, Filippi M. A composite score to predict short-term disease activity in patients with relapsing-remitting MS. Neurology. 2007;69(12):1230-1235. doi:10.1212/01.wnl.0000276940.90309.15
8. Brown FS, Glasmacher SA, Kearns PKA, et al. Systematic review of prediction models in relapsing remitting multiple sclerosis. PLOS ONE. 2020;15(5):e0233575. doi:10.1371/journal.pone.0233575
9. Held U, Heigenhauser L, Shang C, Kappos L, Polman C, Sylvia Lawry Centre for MS Research. Predictors of relapse rate in MS clinical trials. Neurology. 2005;65(11):1769-1773. doi:10.1212/01.wnl.0000187122.71735.1f
10. Liguori M, Meier DS, Hildenbrand P, et al. One year activity on subtraction MRI predicts subsequent 4 year activity and progression in multiple sclerosis. J Neurol Neurosurg Psychiatry. 2011;82(10):1125-1131. doi:10.1136/jnnp.2011.242115
11. Wolff RF, Moons KGM, Riley RD, et al. PROBAST: A Tool to Assess the Risk of Bias and Applicability of Prediction Model Studies. Ann Intern Med. 2019;170(1):51-58. doi:10.7326/M18-1376
12. Royston P, Moons KGM, Altman DG, Vergouwe Y. Prognosis and prognostic research: Developing a prognostic model. BMJ. 2009;338. doi:10.1136/bmj.b604
13. Steyerberg EW. Clinical Prediction Models: A Practical Approach to Development, Validation, and Updating. Springer Science & Business Media; 2008.
14. Steyerberg EW, Moons KGM, Windt DA van der, et al. Prognosis Research Strategy (PROGRESS) 3: Prognostic Model Research. PLOS Med. 2013;10(2):e1001381. doi:10.1371/journal.pmed.1001381
15. Disanto G, Benkert P, Lorscheider J, et al. The Swiss Multiple Sclerosis Cohort-Study (SMSC): A Prospective Swiss Wide Investigation of Key Phases in Disease Evolution and New Treatment Options. PloS One. 2016;11(3):e0152347. doi:10.1371/journal.pone.0152347
16. Transparent reporting of a multivariable prediction model for individual prognosis or diagnosis (TRIPOD): The TRIPOD statement | The EQUATOR Network. Accessed January 20, 2020. https://www.equator-network.org/reporting-guidelines/tripod-statement/

17. Steyerberg EW, Eijkemans MJC, Harrell FE, Habbema JDF. Prognostic modelling with logistic regression analysis: a comparison of selection and estimation methods in small data sets. Stat Med. 2000;19(8):1059-1079. doi: 10.1002/(sici)1097-0258(20000430)19:8<1059::aid-sim412>3.0.co;2-0. PMID: 10790680.

18. Royston P, Sauerbrei W. Multivariable Model - Building: A Pragmatic Approach to Regression Anaylsis based on Fractional Polynomials for Modelling Continuous Variables. Wiley. 2008.

19. Steyerberg EW, Eijkemans MJ, Harrell FE, Habbema JDpubmeddev. Prognostic modeling with logistic regression analysis: in search of a sensible strategy in small data sets. - Med Decis Making. 2001 Jan-Feb;21(1):45-56. doi: 10.1177/0272989X0102100106. PMID: 11206946.

20. Moons KGM, Altman DG, Reitsma JB, et al. Transparent Reporting of a multivariable prediction model for Individual Prognosis or Diagnosis (TRIPOD): explanation and elaboration. Ann Intern Med. 2015;162(1):W1-73. doi:10.7326/M14-0698

21. Riley RD, Snell KI, Ensor J, Burke DL, Harrell FE Jr, Moons KG, Collins GS. Minimum sample size for developing a multivariable prediction model: PART II - binary and time-to-event outcomes. Stat Med. 2019 Mar 30;38(7):1276-1296. doi: 10.1002/sim.7992. Epub 2018 Oct 24. Erratum in: Stat Med. 2019 Dec 30;38(30):5672. PMID: 30357870; PMCID: PMC6519266.

22. Harrell FE. Regression Modelling Strategies: With Applications to Linear Models, Logistic Regression, and Survival Analysis. Springer; 2015.

23. Tibshirani R. Regression Shrinkage and Selection Via the Lasso. J R Stat Soc Ser B Methodol. 1996;58(1):267-288. doi:10.1111/j.2517-6161.1996.tb02080.x

24. O'Hara RB, Sillanpää MJ. A review of Bayesian variable selection methods: what, how and which. Bayesian Anal. 2009;4(1):85-117. doi:10.1214/09-BA403

25. Genkin A, Lewis DD, Madigan D. Large-Scale Bayesian Logistic Regression for Text Categorization. Technometrics. 2007;49(3):291-304. doi:10.1198/004017007000000245

26. Quartagno M, Grund S, Carpenter J. jomo: A Flexible Package for Two-level Joint Modelling Multiple Imputation. R J. 2019;11(2):205. doi:10.32614/RJ-2019-028

27. Carpenter J, Kenward M. Multiple Imputation and its Application. Wiley. 2012. doi: 10.1002/9781119942283

28. Vickers AJ, van Calster B, Steyerberg EW. A simple, step-by-step guide to interpreting decision curve analysis. Diagn Progn Res. 2019;3(1):18. doi:10.1186/s41512-019-0064-7
29. Van Calster B, Wynants L, Verbeek JFM, et al. Reporting and Interpreting Decision Curve Analysis: A Guide for Investigators. Eur Urol. 2018;74(6):796-804. doi:10.1016/j.eururo.2018.08.038
30. Vickers AJ, Elkin EB. Decision Curve Analysis: A Novel Method for Evaluating Prediction Models. Med Decis Making. 2006;26(6):565-574. doi:10.1177/0272989X06295361
31. Zhang Z, Rousson V, Lee W-C, et al. Decision curve analysis: a technical note. Ann Transl Med. 2018;6(15). doi:10.21037/atm.2018.07.02
32. Vickers AJ, Van Calster B, Steyerberg EW. Net benefit approaches to the evaluation of prediction models, molecular markers, and diagnostic tests. BMJ. Published online January 25, 2016:i6. doi:10.1136/bmj.i6
33. Stühler E, Braune S, Lionetto F, et al. Framework for personalized prediction of treatment response in relapsing remitting multiple sclerosis. BMC Med Res Methodol. 2020;20. doi:10.1186/s12874-020-0906-6
34. Pellegrini F, Copetti M, Bovis F, et al. A proof-of-concept application of a novel scoring approach for personalized medicine in multiple sclerosis. Mult Scler Houndmills Basingstoke Engl. Published online May 30, 2019:1352458519849513. doi:10.1177/1352458519849513
35. Kalincik T, Manouchehrinia A, Sobisek L, et al. Towards personalized therapy for multiple sclerosis: prediction of individual treatment response. Brain J Neurol. 2017;140(9):2426-2443. doi:10.1093/brain/awx185
36. Signori A, Schiavetti I, Gallo F, Sormani MP. Subgroups of multiple sclerosis patients with larger treatment benefits: a meta-analysis of randomized trials. Eur J Neurol. 2015;22(6):960-966. doi:10.1111/ene.12690
37. Chalkou K, Steyerberg E, Egger M, Manca A, Pellegrini F, Salanti G. A two-stage prediction model for heterogeneous effects of treatments. Stat Med. 2021 Sep 10;40(20):4362-4375. doi: 10.1002/sim.9034.
38. Lublin FD. Relapses do not matter in relation to long-term disability: no (they do). Mult Scler Houndmills Basingstoke Engl. 2011;17(12):1415-1416. doi:10.1177/1352458511427515

39. Collins GS, de Groot JA, Dutton S, et al. External validation of multivariable prediction models: a systematic review of methodological conduct and reporting. BMC Med Res Methodol. 2014;14:40. doi:10.1186/1471-2288-14-40

## *Appendix*

*Section 1. Summary of pre-existing models on RRMS used in our model*

1. Held et al. [9] aimed to determine the contribution of different possible *prognostic factors* available at baseline to the relapse rate in MS. The authors used 821 patients from the *placebo arms of the Sylvia Lawry Centre* for Multiple Sclerosis Research (SLCMSR) database. The relapse number prior to entry into clinical trials together with disease duration were identified as the best predictors for the relapse rate. The authors validated their model, by splitting the datasets into two samples: the training setting and the validation setting.

2. Kalincik et al. [35] presented an *individualized prediction model* using demographic and clinical predictors in patients with MS. Treatment response was analysed separately for disability progression, disability regression, relapse frequency, conversion to secondary progressive disease, change in the cumulative disease burden, and the probability of treatment discontinuation. They used a *large cohort study*, MSBase, with seven disease-modifying therapies. They validated externally the prediction model in a geographically distinct cohort, the Swedish Multiple Sclerosis Registry. Pre-treatment relapse activity and age were associated with the relapse incidence.

3. Liquori et al. [10] aimed to investigate the *prognostic value* of 1 year subtraction MRI (sMRI) on Change in T2 Lession Volume, Relapse rate, and Change in brain parenchyma fraction. They used 127 patients from a cohort followed in a single centre, the Partners MS Center. They used only MRI and sMRI measures as prognostic factors.

4. Pellegrini et al. [34] developed *a prediction model* to predict treatment response in patients with relapsing-remitting multiple sclerosis, using an individual treatment response score, regressing on a set of baseline predictors. They used two *randomized clinical trials*: CONFIRM and DEFINE studies. The outcome of interest was the annualized relapse rate. The prognostic factors they used are: age, short form-36 mental component summary, short form-36 physical component summary, visual function test 2.5%, prior MS treatment (Yes or No), EDSS, timed 25-foot walk, paced auditory serial addition test (known as PASAT), months since last relapse, number of prior relapses, 9-hole peg test, ethnicity and sex.

5. Signori et al. [36] aimed to examine whether there are *subgroups* of RRMS patients who are more responsive to treatments. 9-Hole Peg Test he collect all published *randomized clinical trials* in RRMS reporting a subgroup analysis of treatment effect. Two main outcomes were studied: the annualized relapse rate and the disability progression. The authors meta-analysed the results of the identified studies to compare the relative treatment effects between subgroups. Age, gadolinium activity and EDSS were identified as the statistical important subgroups regarding the response to treatments for annualized relapse rate.

6. Sormani et al. [7] developed and validated a *prognostic model* to identify RRMS patients with a high risk of experiencing relapses in the short term. They used 539 patients from the placebo arm of a double-blind, *placebo-controlled trial* (CORAL study) of oral glatiramer acetate in RRMS. The validation sample consisted of 117 patients from the placebo arm of a double-blind, placebo-controlled trial of subcutaneous glatiramer acetate in RRMS (European/Canadian Glatiramer Acetate study). The variables included in the final model as independent predictors of relapse occurrence were the number of gadolinium enhanced lesions and the number of previous relapses.

7. Stühler et al. [33] presented a framework for *personalized prediction model* of treatment response based on *real-world data* from the NeuroTransData network for patients diagnosed with RRMS. They examined two outcomes of interest: the number of relapses and the disability progression. They used three different approaches (10-fold cross-validation, leave-one-site-out cross-validation, and excluding a test set) to validate their model. The predictors included for the number of relapses are: age, gender, EDSS, current treatment, previous treatment, disease duration, months since last relapse, number of prior relapses, number of prior therapies, prior second-line therapy (Yes or No), duration of the current treatment, duration of the previous treatment, clinical site.

***Appendix Table 1 TRIPOD checklist was followed for the development and the validation of the prognostic model***

| **Section/Topic** | | **Checklist Item** | **Page** |
|---|---|---|---|
| **Title and abstract** | | | |
| Title | 1 | Identify the study as developing and/or validating a multivariable prediction model, the target population, and the outcome to be predicted. | 1 |
| Abstract | 2 | Provide a summary of objectives, study design, setting, participants, sample size, predictors, outcome, statistical analysis, results, and conclusions. | 2-3 |
| **Introduction** | | | |
| Background and objectives | 3a | Explain the medical context (including whether diagnostic or prognostic) and rationale for developing or validating the multivariable prediction model, including references to existing models. | 3-4 |
| | 3b | Specify the objectives, including whether the study describes the development or validation of the model or both. | 4-5 |
| **Methods** | | | |
| Source of data | 4a | Describe the study design or source of data (e.g., randomized trial, cohort, or registry data), separately for the development and validation data sets, if applicable. | 5 |
| | 4b | Specify the key study dates, including start of accrual; end of accrual; and, if applicable, end of follow-up. | 5 |
| Participants | 5a | Specify key elements of the study setting (e.g., primary care, secondary care, general population) including number and location of centres. | 5-6 |
| | 5b | Describe eligibility criteria for participants. | 5-6 |
| | 5c | Give details of treatments received, if relevant. | 5-6 |
| Outcome | 6a | Clearly define the outcome that is predicted by the prediction model, including how and when assessed. | 4,6 |
| | 6b | Report any actions to blind assessment of the outcome to be predicted. | Not relevant |
| Predictors | 7a | Clearly define all predictors used in developing or validating the multivariable prediction model, including how and when they were measured. | 7, 13, 16, Figure 1 |
| | 7b | Report any actions to blind assessment of predictors for the outcome and other predictors. | Not relevant |
| Sample size | 8 | Explain how the study size was arrived at. | 8, 14 |
| Missing data | 9 | Describe how missing data were handled (e.g., complete-case analysis, single imputation, multiple imputation) with details of any imputation method. | 9-10 |
| Statistical analysis methods | 10a | Describe how predictors were handled in the analyses. | 7-10 |
| | 10b | Specify type of model, all model-building procedures (including any predictor selection), and method for internal validation. | 7-10 |
| | 10d | Specify all measures used to assess model performance and, if relevant, to compare multiple models. | 10-12 |
| Risk groups | 11 | Provide details on how risk groups were created, if done. | Not relevant |
| **Results** | | | |
| Participants | 13a | Describe the flow of participants through the study, including the number of participants with and without the outcome and, if applicable, a summary of the follow-up time. A diagram may be helpful. | Table 1 and Appendix table 2 |
| | 13b | Describe the characteristics of the participants (basic demographics, clinical features, available predictors), including the number of participants with missing data for predictors and outcome. | Table 1 and Appendix table 2 |
| Model development | 14a | Specify the number of participants and outcome events in each analysis. | Table 1 and Appendix table 2 |
| | 14b | If done, report the unadjusted association between each candidate predictor and outcome. | Not relevant |
| Model specification | 15a | Present the full prediction model to allow predictions for individuals (i.e., all regression coefficients, and model intercept or baseline survival at a given time point). | Table 2 |
| | 15b | Explain how to the use the prediction model. | 13, 15 |
| Model performance | 16 | Report performance measures (with CIs) for the prediction model. | 13, Table 2 |
| **Discussion** | | | |
| Limitations | 18 | Discuss any limitations of the study (such as nonrepresentative sample, few events per predictor, missing data). | 16-17 |
| Interpretation | 19b | Give an overall interpretation of the results, considering objectives, limitations, and results from similar studies, and other relevant evidence. | 15-17 |
| Implications | 20 | Discuss the potential clinical use of the model and implications for future research. | 15-17 |
| **Other information** | | | |
| Supplementary information | 21 | Provide information about the availability of supplementary resources, such as study protocol, Web calculator, and data sets. | 12, 13, 15 |
| Funding | 22 | Give the source of funding and the role of the funders for the present study. | 29 |

***Appendix Table 2 Frequency of relapse within two years and frequency of treatment per cycle for patients that were included in one cycle only, patients that were included in two cycles and patients that were included in three cycles. Gender, age, and edss at the beginning of 1st cycle separately for patients with one cycle only, with two cycles and with three cycles. Individuals with one cycle are mainly patients that were recently to the study, whereas individuals with three cycles are those who recruited the study, when SMSC started recruiting.***

| **People with cycles observed** | | **Cycle** | **Relapses** | **Treated** | **Female** | **Age Mean (sd)** | **Edss Mean (sd)** |
|---|---|---|---|---|---|---|---|
| One | **324** | 1 | 62 (19%) | 281 (87%) | 203 (63%) | 39.6 (11.6) | 2.3 (1.5) |
| Two | **405** | 1 | 82 (20%) | 360 (89%%) | 556 (69%) | 40.7 (11.0) | 2.3 (1.3) |
| | | 2 | 50 (12%) | 398 (98%) | | | |
| Three | **206** | 1 | 47 (23%) | 184 (89%) | 450 (73%) | 42.6 (10.8) | 2.5 (1.4) |
| | | 2 | 44 (22%) | 196 (95%) | | | |
| | | 3 | 17 (8%) | 199 (97%) | | | |

***Appendix Table 3 Characteristics of the studies used to inform the developed prognostic model***

| **Study** | **Design** | **Sample size (EPV)** | **Outcome** | **Missing data** | **Discrimination Calibration Validation** | **Presentation of the model** | **Model** |
|---|---|---|---|---|---|---|---|
| **PHeld et al.** [9] | Placebo arms of clinical trials Multicentre | n=821 (NA) | Relapse rate | Complete case analysis | **Discrimination:** C statistic NA for continuous outcome **Calibration:** absent. **Internal validation:** split-sample | Table | Prognostic model |
| **Kalincik et al.** [35] | Cohort study | n=8513 | Treatment response for relapse frequency | Mentioned: Values of the principal components can be estimated even for patients with incomplete data | Accuracy and internal validity reported (moderate for relapse rate at 2 years) **Internal validation:** in a separate, non-overlapping | Table of principal components | Prediction model |
| | | | | | MSBase cohort. **External validation:** 2945 patients from the Swedish Multiple Sclerosis Registry | | |
| **Liquori et al.** [10] | Cohort (R) Single centre | n=127 (NA) | Relapse rate | Complete case analysis | Overall performance reported R2 **Calibration:** absent **Validation:** absent | NA | Prognostic model |
| **Pellegrini et al.** [34] | RCT | n=2099 | Treatment response to annualized relapse rate | Complete case analysis | Performance measure of area under the AD(c) curve shown in graph **Calibration** absent **Internal validation**: Splitting of the training dataset into two subsets (50%/50%) **External validation:** Independent RCT | Table | Prediction model |
| **Signori et al.** [36] | All published randomized clinical trials in RRMS reporting a subgroup analysis | Six trials 6693 RRMS patients | Treatment response to annualized relapse rate | Not relevant | Not relevant | Not relevant | Subgroups responsive to treatments |
| **Sormani et al.** [7] | Placebo arm of RCT Multicentre | n=539 (Insufficient data reported) | Number of relapses at 9 months | Complete case analysis | **Discrimination:** absent **Calibration:** absent **Internal validation:** absent | Mathematical formula | Prognostic model |
| **Stühler et al.** [33] | Real-world data | n=25000 | Treatment response to number of relapses | Complete case analysis | **Discrimination:** c-statistic (0.65) **Calibration:** Calibration plot **Internal validation :** 1)10-fold cross-validation, 2)leave-one-site-out cross-validation, and | Table | Prediction model |

3)excluding
test set

***Appendix Table 4 The estimation of all parameters in the complete dataset and in each one of the imputed datasets***

| Parameters | CC | ID 1 | ID 2 | ID 3 | ID 4 | ID 5 | ID 6 | ID 7 | ID 8 | ID 9 | ID 10 |
|---|---|---|---|---|---|---|---|---|---|---|---|
| **Intercept** | -2.35 (-3.99, 0.82) | -1.89 (-3.14, -0.52) | -1.95 (-3.23, -0.51) | -1.99 (-3.23, -0.67) | -1.83 (-3.06, -0.44) | -1.82 (-3.1, -0.42) | -1.86 (-3.19, -0.48) | -1.94 (-3.23, -0.56) | -1.9 (-3.16, -0.49) | -1.83 (-3.04, -0.44) | -1.83 (-2.98, -0.53) |
| **Age** | -0.03 (-0.05, 0.02) | -0.04 (-0.05, -0.02) | -0.04 (-0.05, -0.02) | -0.03 (-0.05, -0.02) | -0.04 (-0.05, -0.02) | -0.04 (-0.05, -0.02) | -0.04 (-0.05, -0.02) | -0.04 (-0.05, -0.02) | -0.03 (-0.05, -0.02) | -0.03 (-0.05, -0.02) | -0.03 (-0.05, -0.02) |
| **Disease duration** | 0.32 (-0.09, 0.89) | 0.32 (-0.04, 0.78) | 0.27 (-0.07, 0.7) | 0.34 (-0.03, 0.78) | 0.41 (-0.01, 0.91) | 0.42 (-0.01, 0.92) | 0.34 (-0.04, 0.81) | 0.36 (-0.03, 0.85) | 0.29 (-0.06, 0.72) | 0.29 (-0.06, 0.72) | 0.18 (-0.09, 0.5) |
| **Edss** | 0.1 (-0.02, 0.22) | 0.12 (0.02, 0.23) | 0.12 (0.01, 0.23) | 0.13 (0.02, 0.23) | 0.12 (0.01, 0.22) | 0.12 (0.01, 0.22) | 0.12 (0.02, 0.22) | 0.12 (0.02, 0.23) | 0.13 (0.02, 0.23) | 0.13 (0.02, 0.23) | 0.14 (0.04, 0.24) |
| **Number of Gd enhanced lesions (>0 vs 0)** | 0.04 (-0.35, 0.49) | -0.01 (-0.38, 0.33) | 0.1 (-0.24, 0.51) | 0.14 (-0.18, 0.56) | 0 (-0.38, 0.35) | -0.04 (-0.4, 0.29) | 0.01 (-0.35, 0.38) | 0.06 (-0.27, 0.44) | 0.05 (-0.27, 0.41) | 0.02 (-0.31, 0.36) | 0.01 (-0.3, 0.35) |
| **Prior Relapses (1 vs 0)** | 0.02 (-0.32, 0.38) | -0.06 (-0.39, 0.22) | -0.07 (-0.42, 0.21) | -0.08 (-0.41, 0.2) | -0.09 (-0.45, 0.19) | -0.09 (-0.43, 0.21) | -0.08 (-0.42, 0.22) | -0.07 (-0.41, 0.22) | -0.07 (-0.39, 0.21) | -0.08 (-0.41, 0.2) | -0.07 (-0.38, 0.19) |
| **Prior Relapses (≥2 vs 0)** | 0.08 (-0.26, 0.49) | 0.14 (-0.17, 0.53) | 0.11 (-0.24, 0.48) | 0.11 (-0.2, 0.47) | 0.11 (-0.21, 0.47) | 0.12 (-0.21, 0.52) | 0.13 (-0.2, 0.51) | 0.12 (-0.19, 0.49) | 0.12 (-0.19, 0.5) | 0.12 (-0.19, 0.48) | 0.13 (-0.16, 0.48) |
| **Months since last relapse** | -0.36 (-0.63, -0.1) | -0.47 (-0.7, -0.25) | -0.45 (-0.72, -0.2) | -0.5 (-0.72, -0.28) | -0.5 (-0.74, -0.28) | -0.5 (-0.74, -0.28) | -0.49 (-0.72, -0.26) | -0.48 (-0.72, -0.26) | -0.48 (-0.71, -0.26) | -0.48 (-0.72, -0.26) | -0.46 (-0.67, -0.25) |
| **Treatment naïve (Yes vs No)** | 0.15 (-0.21, 0.65) | 0.08 (-0.21, 0.43) | 0.07 (-0.24, 0.41) | 0.07 (-0.21, 0.42) | 0.1 (-0.2, 0.46) | 0.1 (-0.2 , 0.47 ) | 0.08 (-0.21, 0.43) | 0.09 (-0.21, 0.44) | 0.07 (-0.23, 0.41) | 0.07 (-0.22, 0.42) | 0.06 (-0.22, 0.4) |
| **Gender (Female vs Male)** | 0.43 (0.05, 0.83) | 0.26 (0, 0.56) | 0.24 (-0.02, 0.56) | 0.27 (0 , 0.57) | 0.25 (-0.01, 0.54) | 0.26 (-0.01, 0.56) | 0.25 (0 , 0.56) | 0.25 (-0.01, 0.55) | 0.25 (-0.01, 0.56) | 0.24 (-0.01, 0.54) | 0.24 (-0.01, 0.54) |
| **Currently on treatment (Yes vs No)** | -0.14 (-0.68, 0.28) | -0.21 (-0.73, 0.16) | -0.23 (-0.81, 0.15) | -0.22 (-0.73, 0.16) | -0.22 (-0.75, 0.16) | -0.23 (-0.74, 0.14) | -0.22 (-0.73, 0.17) | -0.21 (-0.73, 0.17) | -0.21 (-0.75, 0.15) | -0.22 (-0.76, 0.15) | -0.22 (-0.73, 0.13) |
| $\boldsymbol{\rho}$ | 0.62 (-0.05, 0.99) | 0.73 (0.23, 0.99) | 0.42 (-0.06, 0.95) | 0.64 (-0.09, 0.99) | 0.31 (-0.1, 0.97) | 0.43 (-0.05, 0.97) | 0.56 (0.01 , 1) | 0.26 (-0.1, 0.85) | 0.48 (-0.02, 0.99) | 0.58 (-0.05, 0.98) | 0.6 (-0.02, 1) |
| $\boldsymbol{\sigma}$ | 0 (0, 0.02) | 0 (0, 0.02) | 0.03 (0, 0.14) | 0 (0 , 0) | 0.02 (0, 0.11) | 0.01 (0 , 0.04) | 0.02 (0 , 0.09 ) | 0.01 (0, 0.02) | 0.01 (0, 0.03) | 0.01 (0, 0.02) | 0 (0, 0.01) |

CC: Complete cases dataset; ID: Imputed Dataset; Gd: Gadolinium.